\documentclass[12pt,showpacs]{revtex4}
\usepackage{graphicx}
\begin{document}

\title{CONTROLLING THE SPIN POLARIZATION OF THE ELECTRON CURRENT IN A SEMIMAGNETIC RESONANT-TUNNELING DIODE}
\author{N.N. Beletskii$^{1,2}$, G.P. Berman$^2$, and S.A. Borysenko$^1$}
\affiliation{$^1$Usikov Institute of Radiophysics and Electronics
of the National Academy of Sciences of Ukraine, 12, Acad. Proskura
Str.,61085 Kharkov, Ukraine}

\affiliation{$^2$Theoretical Division, MS B213, Los Alamos
National Laboratory, Los Alamos, New Mexico 87545}

\vspace{3mm}

\begin{abstract}
The spin filtering effect of the electron current in a
double-barrier resonant-tunneling diode (RTD) consisting of
Zn$_{1-x}$Mn$_{x}$Se semimagnetic layers has been studied
theoretically. The influence of the distribution of the magnesium
ions on the coefficient of the spin polarization of the electron
current has been investigated. The dependence of the spin
filtering degree of the electron current on the external magnetic
field and the bias voltage has been obtained. The effect of the
total spin polarization of the electron current has been
predicted. This effect is characterized by total suppression of
the spin-up component of electron current, that takes place when
the Fermi level coincides with the lowest Landau level for spin-up
electrons in the RTD semimagnetic emitter.
\end{abstract}

 \maketitle

\section*{Introduction}
Spin-polarized ballistic electron transport in resonant-tunneling
semimagnetic semiconductor nanostructures attracts considerable
attention of the researchers developing the fundamentals of
spintronics
\cite{Wessel,Chitta,Egues1,Egues2,Guo1,Guo2,Guo3,Slobodskyy}. This
transport is also associated with the search for effective sources
of the spin-polarized current which can be controlled using a
constant magnetic field $\mathbf{B}$ as well as by means of a bias
voltage $V_a$. Resonant-tunneling semimagnetic nanostructures are
characterized by the high degree of the current spin polarization
due to the $sp-d$ exchange interaction between the conduction
electrons and localized electrons of the magnetic ions belonging
to the semimagnetic semiconductors \cite {Furd,Goede,Eunsoon}. In
a magnetic field $\mathbf{B}$, this interaction gives rise to the
giant Zeeman splitting of the electron energy levels. As a result,
the electrons with spins oriented along $\mathbf{B}$ (spin-up
electrons) and against $\mathbf{B}$ (spin-down electrons) move in
different potential fields and have different transmission
coefficients through the resonant-tunneling semimagnetic
semiconductor nanostructures. Therefore, spin filtering of the
electron current occurs even in moderate  magnetic fields, and the
electrons with a certain spin direction dominate in the current.
The presence of the spin filtering of the electron current can be
detected by its injection into a light-emitting diode and by the
measurement of the electromagnetic radiation of the circular
polarization \cite{Fiederling,Jonker}.

The idea of using semimagnetic semiconductors for spin filtering
of the electron current has been proposed in \cite{Egues1}. It was
shown that the electron current flowing through a semimagnetic
semiconductor layer in a constant magnetic field of 2-4 T displays
a high degree of spin polarization. In \cite{Guo1}, the
dependences of the coefficient of current spin polarization on the
thickness of the semimagnetic layer and the bias voltage have been
investigated. In  \cite{Guo2,Guo3}, the results of papers
\cite{Egues1,Guo1} have been summarized for the case of a
nanostructure consisting of two semimagnetic semiconductor layers
separated by a non-magnetic layer. In these papers along with the
study of voltage-current characteristics of the nanostructure, the
influence of the thicknesses of semimagnetic layers \cite{Guo2}
and operating temperatures \cite{Guo3} on the value of the
coefficient of the current spin polarization has been
investigated.

Later, it was shown that the degree of the current spin
polarization can be enhanced if the resonant-tunneling
nanostructure has semimagnetic contacts \cite{Egues2}. This is
related to the fact that the conduction band edge of a
semimagnetic emitter in the magnetic field $\mathbf{B}$ is
spin-dependent. In this case, the number of spin-down electrons in
the emitter exceeds the number of spin-up electrons. As a result,
spin-down electrons  play the determining role in the current
flowing through the resonant-tunneling nanostructure with
semimagnetic contacts. Thus, the spin-dependent shift of the
conduction band edge of the semimagnetic emitter and the
spin-dependent electron transmission through semimagnetic layers
lead to a significant increase in the coefficient of current spin
polarization in fully semimagnetic resonant-tunneling
nanostructures.

In this paper new results are presented on the theory of the
effect of the electron current spin filtering in a double-barrier
resonant-tunneling diode (RTD) based on a Zn$_{1-x}$Mn$_{x}$Se
semimagnetic semiconductor. The choice of this semimagnetic
semiconductor is related to the presence of an RTD in which the
emitter, collector, and quantum well consist of this semiconductor
material \cite{Slobodskyy}. In contrast to the paper
\cite{Egues2}, we assume that all RTD layers are semimagnetic.
Moreover, in our paper the value of the electron current density
and the coefficient of current spin polarization are determined
taking into account the influence of the bias voltage $V_a$ on the
coefficient of the electron transmission through the RTD.

The dependencies of the electron current density and the
coefficient of the current spin polarization on the constant
magnetic field $\mathbf{B}$ as well as on a bias voltage $V_a$ are
studied for different  spatial distributions of magnetic ions in
the RTD and for different values of the Fermi level in the RTD
emitter. The occurrence of the  total polarization of the electron
current has been predicted. Total polarization takes place when
the Fermi level coincides with the lowest Landau level for spin-up
electrons in the RTD semimagnetic emitter.

\section*{Theoretical model}

We assume that the RTD (including its emitter and collector)
consists of Zn$_{1-{x_j}}$Mn$_{x_j}$Se layers with different Mn
concentrations $x_{j}=\{x_1,x_2,x_3,x_4,x_5\}$ (Fig.1a). The
region $z<z_1$ is an RTD emitter and the region $z>z_4$ is a RTD
collector. We assume that the emitter and collector are $n$-doped.
The external magnetic field $\mathbf{B}$ is directed along the $z$
axis. The bottoms of the conduction bands for the spin-down and
spin-up electrons are shown by the solid and dashed lines
correspondingly in Fig.1b. The values of $L_i=z_{i+1}-z_i$
($i=1,2,3$) are the thicknesses of two potential barriers ($L_1$
and $L_3$ ) and potential well ($L_2$) of the RTD. The value $E_F$
is the Fermi level in the emitter and collector.

\begin{figure}
\mbox{
\includegraphics[width=16cm,height=14cm]{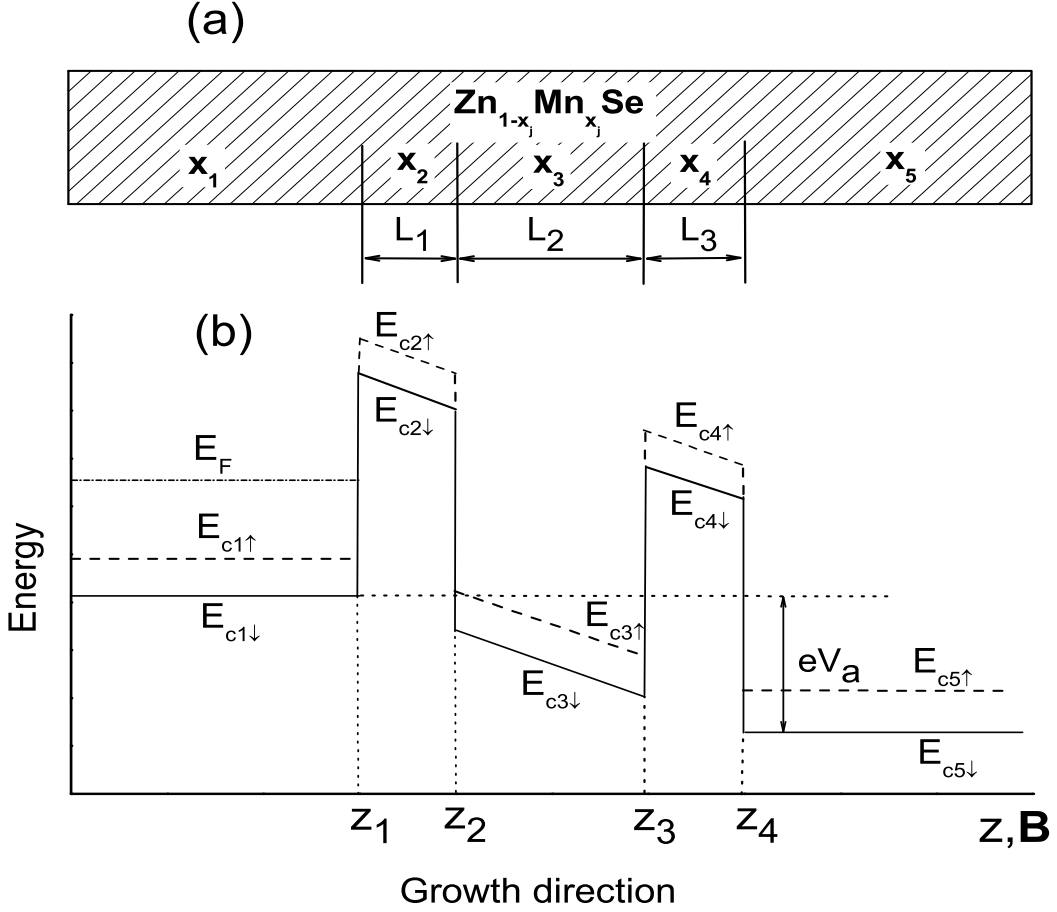}}
\vspace{-5mm} \caption{(a) Zn$_{1-x}$Mn$_{x}$Se double-barrier
resonant-tunnelling semimagnetic
 nanostructure (RTD)  and (b) its spin-dependent conduction band profile in the nonzero magnetic
 field.}
\label{fig:1}
\end{figure}

As is well known, the band gap of semimagnetic semiconductors
depends on the Mn concentration
\cite{Furd,Klar,Bylsma,Superlattice2}. Therefore at the boundary
between two semimagnetic semiconductors with different Mn
concentrations, an offset of the band gap takes place. In this
case, one part of this offset falls at the conduction band offset
and the other one falls at the valence band offset \cite{Klar}. At
low temperatures the band gap $E_{gj}$ of the semimagnetic
semiconductors depends slightly on $x_{j}$ in the range $x_3 <
0.065$ \cite{Furd,Klar,Bylsma,Superlattice2}. Therefore, to obtain
the dependence of the conduction band edge $E_{cj}(x_{j})$ of the
semimagnetic semiconductor,  we use the following empirical
formula which describes the experimental dependencies in
\cite{Klar}:

\begin{equation}
  E_{cj}(x_{j})=\left\{%
\begin{array}{ll}
    E_{g}(0), &  x_{j}<0.065, \\
    E_{0}+(1-{\rm VBO})x_{j}\Delta E_{g} &  x_{j} > 0.065. \\
\end{array}%
\right.
  \label{eq1}
\end{equation}
Here $ E_{g}(0)=2.822$ {\rm eV} is the band gap of ZnSe; {\rm VBO}
is the valence band offset; $\Delta E_{g}=0.4141$ {\rm eV},
$E_{0}$ is the fitting parameter (for each value of {\rm VBO}, it
is determined in such a way that at the point $x_j=0.065$ the
function $E_{cj}(x_{j})$ is continuous).

In the external magnetic field $\mathbf{B}$, the conduction band
edge of the semimagnetic semiconductor is spin-dependent due to
the effect of the giant Zeeman splitting of the electron energy
levels \cite{Furd,Goede}. The value of the spin-dependent shift
$\epsilon_{j\sigma_z}(B)$ of the conduction band edge of the
semimagnetic semiconductor is equal to the value of the energy of
the $sp$-$d$ exchange interaction between the conduction electrons
and localized electrons of the magnetic Mn ions
\begin{equation}
\epsilon_{j\sigma_z}(B)=-\sigma_z x_{j}^{eff} N_0\alpha\langle
S_{zj} \rangle,
\label{eq2}
\end{equation}
where $\sigma_z=\pm 1/2$ (or $\uparrow,\downarrow$) is the spin
quantum number; $x_{j}^{eff}=x_j(1-x_j)^{12}$ is the effective
concentration of Mn ions \cite{Egues1,Egues2,Guo1,Guo2,Guo3};
$N_0\alpha$ is the $sp$-$d$ exchange constant for conduction
electrons; and $\langle S_{zj} \rangle$ is the thermal average of
the Mn spin component along the magnetic field $\mathbf{B}$
\begin{equation}
\langle S_{zj} \rangle=-S B_S( g_{Mn}\mu_{B}SB/kT_{j}^{eff}).
\label{eq3}
\end{equation}
Here $B_S$ is the modified Brillouin function for the total spin
quantum number of Mn ions; $S=5/2$; $g_{Mn}=2$ is $g$-factor of
the spectroscopic splitting for Mn-$d$-electrons; $\mu_{B}$ is the
Bohr magneton; $T_{j}^{eff}=T+T_{j}^{AF}$ is the effective
temperature; $T$~ is  the lattice temperature of semimagnetic
semiconductors; and $T_{j}^{AF}$ is the phenomenological
parameter. The parameters, $x_{j}^{eff}$ and $T_{j}^{AF}$, are
required by the necessity to take into account the
antiferromagnetic interaction between the Mn ions.

Thus, the conduction band edge of semimagnetic semiconductors
$E_{cj\sigma_z}$ in the magnetic field $\mathbf{B}$ is determined
by the following formula
\begin{equation}
E_{cj\sigma_z}=E_{cj}+\epsilon_{j\sigma_z}(B).
 \label{eq4}
\end{equation}

We consider sufficiently high magnetic fields for which the Landau
quantization of transverse motion of electrons is important. Then
the electron energy in each layer of the considered RTD has the
following form:
\begin{equation}
             E_{j\sigma_z}=E_{cj\sigma_z}+
             (l+\frac{1}{2})\hbar\omega_c+\sigma_z
             g^* \mu_B B+E_z.
             \label{eq5}
 \end{equation}
Here $l=0,1,2,\ldots$ is the Landau level quantum number;
$\omega_c=eB/cm^*$ is
 the electron cyclotron frequency; $E_z=\hbar^2k_z^2/2m^*$ is
 the electron energy connected with their  motion along the RTD ($k_z$ is the electron wave vector along
 $z$ direction);
  $m^*$~ is the effective electron mass
 (we assume a single electron mass throughout all RTD layers); and
 $g^*$ is the zone electron $g$-factor.

Taking into account expression (\ref{eq4}), the electron energy in
each RTD layer can be written in the following form
\begin{equation}
             E_{j\sigma_z}=E_{gj}+
             (l+\frac{1}{2})\hbar\omega_c+
             \sigma_z g_{j}^{eff} \mu_B B+E_z,
             \label{eq6}
\end{equation}
where
\begin{equation}
             g_{j}^{eff}=g^*+x_{j}^{eff} N_0\alpha S B_S( g_{Mn}\mu_{B}SB/kT_{j}^{eff})
             /\mu_B B.
             \label{eq7}
\end{equation}
The average current density through the RTD created by electrons
with $\sigma_z$ polarization in the magnetic field $B$ at the
finite temperature $T$ is determined by the following expression
\cite{Guo1,Guo2,Guo3}:
\begin{eqnarray}
        J_{\sigma_z}&=&J_0
             B\sum_{l=0}^\infty\int_{0}^\infty T_{\sigma_z}(E_z,B,V_a)
             \{f[E_z+(l+\frac{1}{2})\hbar\omega_{c}+\sigma_z
             g_{1}^{eff} \mu_B B]- \nonumber \\
             & & f[E_z+(l+\frac{1}{2})\hbar\omega_{c} +eV_a +\sigma_z
             g_{5}^{eff} \mu_B B]\}dE_{z},
             \label{eq8}
\end{eqnarray}
where $T_{\sigma_z}(E_z,B,V_a)$ is the electron transmission
coefficient through the RTD; $J_0=e^{2}/h^{2}c$; and
$f(E)=1/(1+\exp((E-E_F)/kT))$ is the Fermi function.

The total current density  $J_t$ through the RTD is
$J_{\uparrow}+J_{\downarrow}$ and the coefficient of the current
spin polarization $P$ is
\begin{equation}
             P=\frac{J_{\downarrow}-J_{\uparrow}}{J_{\downarrow}+J_{\uparrow}}.
             \label{eq9}
\end{equation}
To find $T_{\sigma_z}(E_z,B,V_a)$ we use the Airy's-function-based
transfer-matrix method \cite{Airy}. This allows us to calculate
$J_{\sigma_z}$ numerically for arbitrary values of $V_a$. In the
following we use these specific values of the RTD parameters:
$m^*=0.16m_0$ ($m_0$ is the free electron  mass), $g^*=1.1$;
$N_0\alpha=0.26$~eV; $T$=4.2 K; $T_{eff}^j$= 2 K; $L_1=L_3=5$ nm;
and $L_2=9$ nm. Note that the thicknesses of the quantum well and
two barriers of the RTD correspond to the physical semimagnetic
RTD with non-magnetic barriers, whose properties were investigated
experimentally in \cite{Slobodskyy}.

\section*{Numerical results and discussion}

The spin-filtering effect of the electron current becomes most
clearly apparent when the energy of the $sp$-$d$ exchange
interaction is maximal. Considering this energy as a function of
$x_j$, it is easy to show from formula (\ref{eq2}) that it is
maximal at $x_j=x_{m}=1/13 \approx 0.077$. Later on we will
consider the case when the Mn concentration in the emitter and
collector of the RTD is equal to this value, that is
$x_1=x_5=x_{m}$. This allows us to obtain the maximal value of the
spin-dependent shift of the conduction band edge of the emitter
and collector. To create the potential profile inherent in
double-barrier RTDs, it is required that the concentration of Mn
ions in the two barriers ($x_2$ and $x_4$) is larger than in the
emitter ($x_1$), collector ($x_5$), and the potential well
($x_3$). We assume that $x_2=x_4=0.25$, and $x_3$ is changed from
$x_3=0$ to $x_3=x_{m}$.

\begin{figure}
\mbox{
\includegraphics[width=16cm,height=12cm]{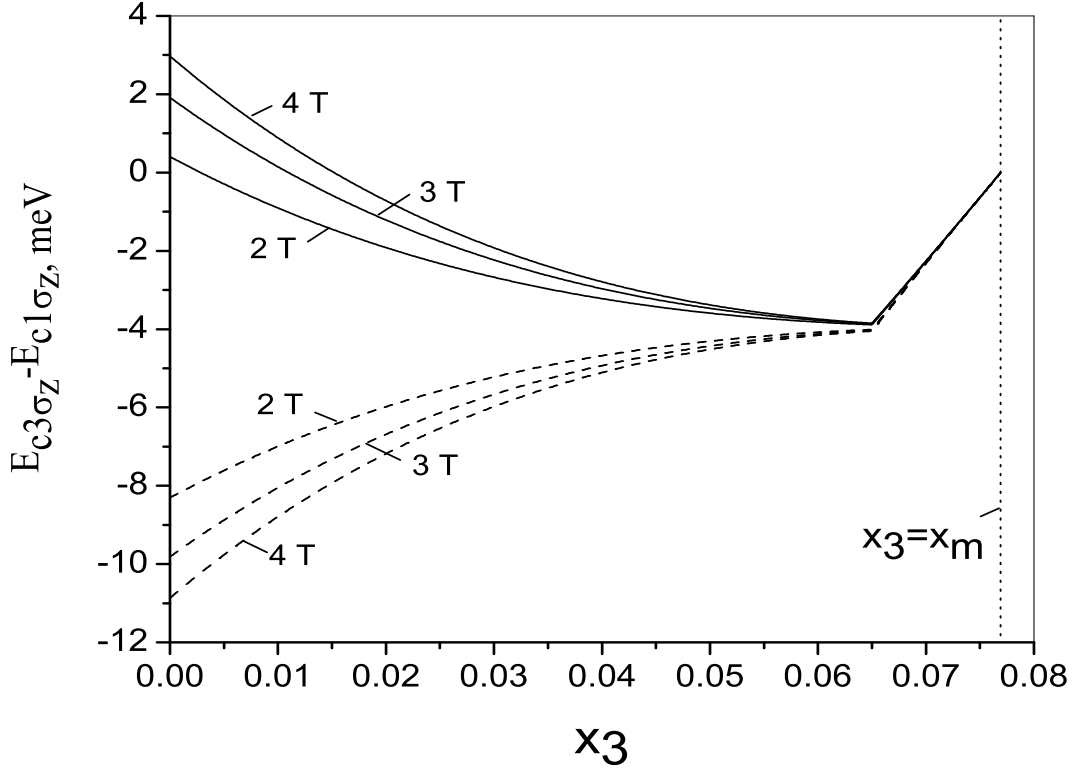}}
\vspace{-5mm} \caption{Dependence of the conduction-band edge of
the RTD quantum well on the Mn-ion concentration $x_3$ for
spin-down (solid lines) and spin-up (dashed lines) electrons for
 $B=2,3,4$~T.} \label{fig:2}
\end{figure}

\subsection {The spin-dependent RTD conduction band profile at the zero
bias voltage}

Fig.2 shows the dependence of the spin-dependent conduction band
edge of the RTD quantum well on the Mn concentration $x_3$ for
three values of $B=2,3,4$ T. In this case, we choose the zero of
the energy to be at the conduction band edge of the RTD emitter
(the solid lines correspond to the spin-down electrons and the
dashed lines correspond to the spin-up electrons). It is seen from
Fig.2 that with increasing $B$, the difference in the position of
the conduction band edges of the RTD quantum well for the spin-up
and spin-down electrons increases. At a fixed value of $B$, the
largest difference in the position of the spin-dependent
conduction band edges takes place, and hence the largest spin
splitting of the electron levels in the RTD quantum well occurs,
at $x_3=0$. For this reason we begin our study of the value of the
total RTD current density $J_t$ and the value of the coefficient
of the RTD current spin polarization $P$ with this case.

\begin{figure}
\mbox{
\includegraphics[width=16cm,height=14cm]{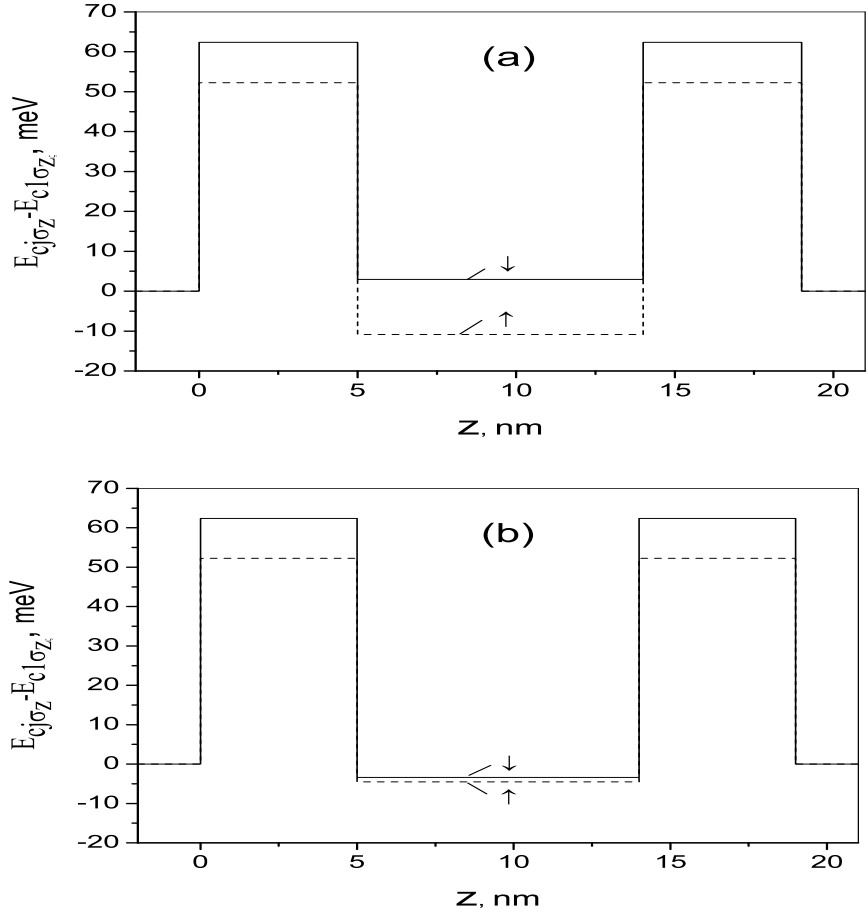}}
\vspace{-5mm} \caption{The zero bias voltage RTD potential profile
for spin-down electrons (solid lines) and for spin-up electrons
(dashed lines) for (a) $x_3=0.0$ and (b) $x_3=0.05$ at $B=4$~T.}
\label{fig:3}
\end{figure}

Fig.3 shows the zero bias voltage RTD potential profile (the
energy is measured from the conduction band edge of the emitter,
$z_1=0$) for spin-down electrons (solid lines) and for spin-up
electrons (dashed lines) for (a) $x_3=0.0$ and (b) $x_3=0.05$ at
$B=4$~T. One can see that for spin-up electrons the barriers are
smaller and the quantum well is deeper than for spin-down
electrons. Consequently, the energy levels in the quantum well lie
deeper for spin-up electrons than for the spin-down electrons. It
is obvious that with decreasing $x_3$, the difference in the
potential profile for spin-up and spin-down electrons increases,
and the effect of electron current spin filtering becomes more
apparent.

\subsection{Magnetic field dependencies of the RTD current spin polarization}

 In Fig.4  the dependencies of $J_{\uparrow}(V_a)$,
$J_{\downarrow}(V_a)$, $J_t(V_a)$ (the left axis of ordinates) and
$P(V_a)$ (the right axis of ordinates) are shown at (a) $B=2$~T
and (b) $B=4$~T for $x_3=0.0$ and $E_F=10$~meV. It is seen from
these figures that there are two  current density peaks in the
curves $J_{\uparrow}(V_a)$ and $J_{\downarrow}(V_a)$. With
increasing $B$, the values of the peaks of $J_{\downarrow}(V_a)$
increase and for $J_{\uparrow}(V_a)$ they decrease. In this case,
the values of the peaks of the total current density $J_t(V_a)$
increase when $B$ increases. The dependencies of $P(V_a)$ are
non-monotone functions and the values of the peaks of $P$ rise
with increasing $B$ as well. The low-voltage range is of interest,
in which $P\approx 1$ for rhe relatively small value of $B=2$ T.
In Fig.4, the presence of two peaks of the current densities
$J_{\downarrow}(V_a)$ and $J_{\uparrow}(V_a)$ corresponds to the
two lowest resonant spin splitting electron energy levels in the
RTD quantum well. In this case, as the bias voltage increases the
resonant electron transmission takes place, from the beginning,
for the first lowest electron energy level in the quantum well and
then for the second electron energy level. Note that the shape of
the first peak in $J_t(V_a)$ has interesting features such as at
$B=2$~T the current density peak is split and at $B=4$~T there are
kinks. This is due to both the presence of the spin splitting of
the electron energy levels in the quantum well and the
quantization of the transverse electron motion (the presence of
the Landau levels).

\begin{figure}
\mbox{
\includegraphics[width=16cm,height=16cm]{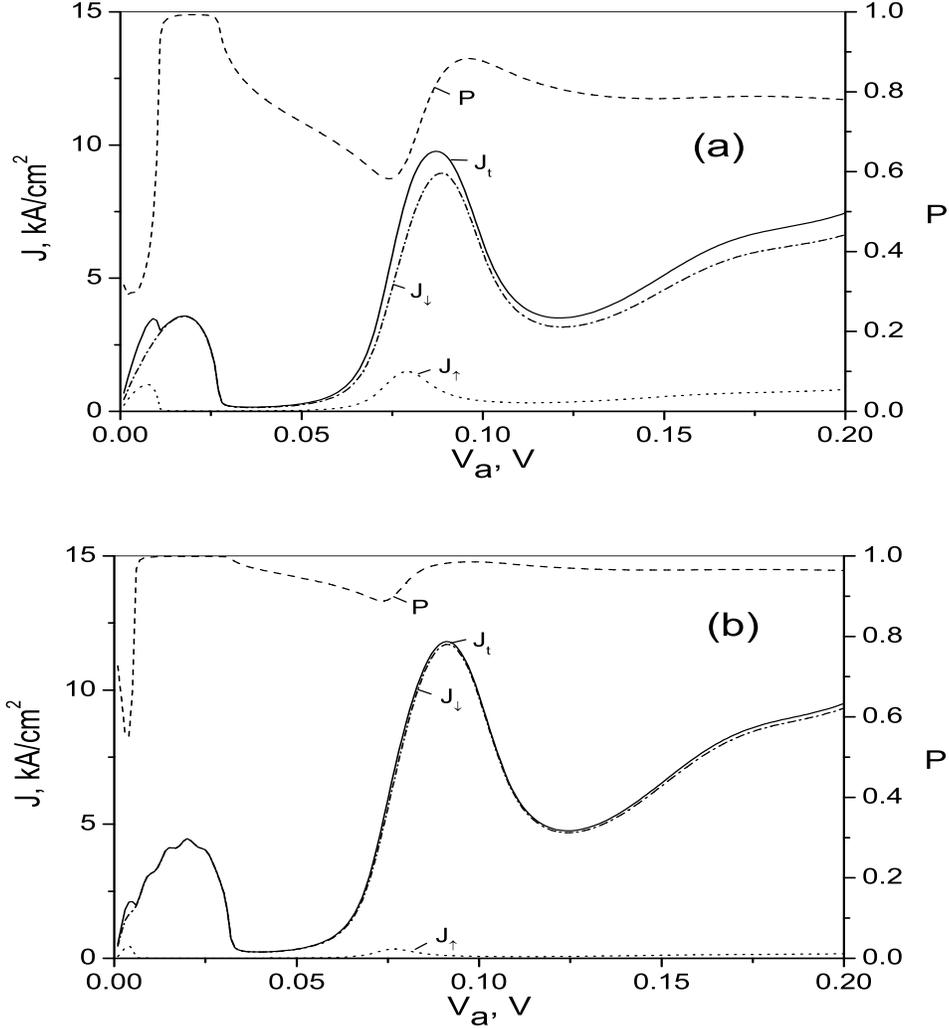}}
\vspace{-5mm} \caption{$J_{\uparrow}(V_a)$, $J_{\downarrow}(V_a)$,
$J_t(V_a)$ (the left axis of ordinates) and $P(V_a)$ (the right
axis of ordinates) at (a) $B=2$~T and (b) $B=4$~T for $x_3=0.0$,
$E_F=10$~meV.} \label{fig:4}
\end{figure}

A note should be made concerning the physical phenomena
determining the shape of the above-mentioned dependencies
$J_t(V_a)$ and $P(V_a)$. For this reason we plot
$T_{\downarrow}(E_z)$ (Fig.5a) and $T_{\uparrow}(E_z)$ (Fig.5b)
for the different values of the voltage bias $V_a$ for $B=4$ T and
$x_3=0.0$ (the numbers next to the curves show the corresponding
values of $V_a$ in volts).

\begin{figure}
\mbox{
\includegraphics[width=16cm,height=16cm]{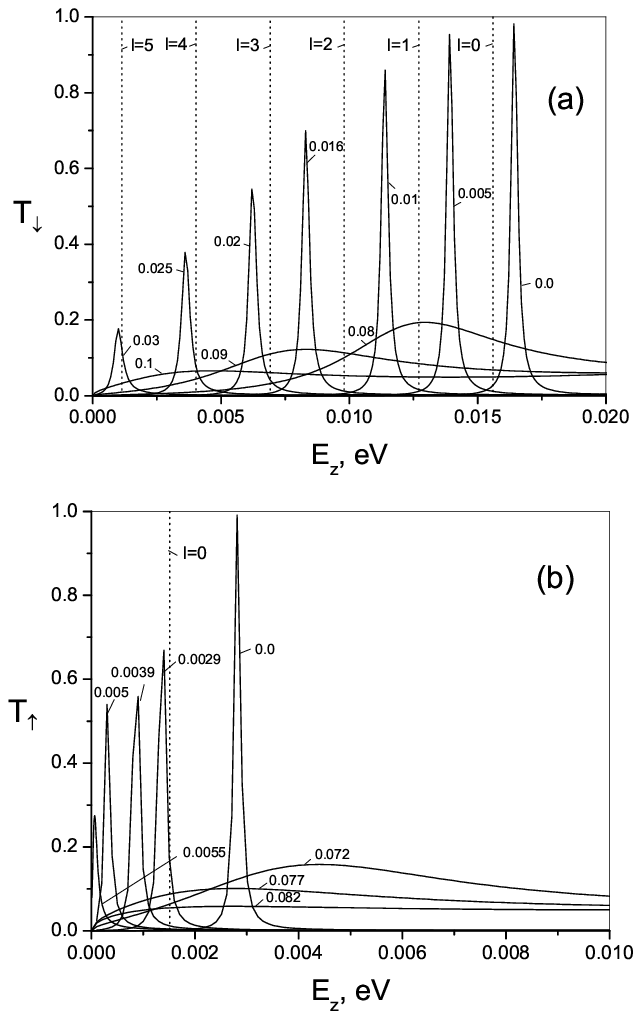}}
\vspace{-5mm} \caption{(a) $T_{\downarrow}(E_z)$ and (b)
$T_{\uparrow}(E_z)$ for the different values of the voltage bias
$V_a$ at $B=4$ T and $x_3=0.0$ (the numbers next to the curves
show the corresponding values of $V_a$ in volts).} \label{fig:5}
\end{figure}

There are resonant peaks with unit peak-value in
$T_{\downarrow}(E_z)$ and $T_{\uparrow}(E_z)$ for $V_a=0$ (in view
of the chosen scale in Fig.5, these curves show only the region of
the first resonant peak both for $T_{\downarrow}(E_z)$ and for
$T_{\uparrow}(E_z)$). Due to the fact that the depth of the
potential well depends significantly on the electron spin, the
resonant peaks of $T_{\downarrow}(E_z)$ and $T_{\uparrow}(E_z)$
strongly differ in location. With increasing $V_a$ the resonant
peaks of $T_{\downarrow}(E_z)$ and $T_{\uparrow}(E_z)$ shift in
the low-energy region, and their peak values decrease. In Fig.5
and Fig.6 the dotted lines show the values of
\begin{equation}
  E_{zm}^{\sigma_z}(l)=E_F-\frac{1}{2}\hbar\omega_c-\sigma_z
g_{1}^{eff} \mu_B B,
             \label{eq10}
\end{equation}
which are the maximal values of the longitudinal electron energy
$E_z$ for each Landau level $l$. The electrons located at Landau
level $l$ pass through the RTD when $E_{zm}^{\sigma_z}(l)>0$. For
$\sigma_z=1/2$, this condition is fulfilled only for $l=0$, but at
$\sigma_z=-1/2$ it holds for $l=0,\ldots,5$. With increasing $V_a$
the current through the RTD occurs as soon as the first resonant
peak of $T_{\sigma_z}$ intersects the line
$E_z=E_{zm}^{\sigma_z}(0)$ for the Landau level, $l=0$. For
spin-down electrons this takes place at $V_a=0.002$ V and for the
spin-up electrons it occurs at $V_a=0.0026$~V. It is seen from
Fig.5a that with increasing $V_a$ the resonant peak of
$T_{\downarrow}(E_z)$ shifts towards the low-energy region. In
this case, the resonant peak decreases in magnitude and
successively intersects the lines $E_z=E_{zm}^{\sigma_z}(l)$. At
each intersection, the current density $J_{\downarrow}$ increases
at the expense of the electrons located at the corresponding
Landau levels $l$, and a kink in $J_{\downarrow}(V_a)$ occurs. On
the other hand, the decrease in the magnitude of the resonant peak
of $T_{\downarrow}$ leads to a decrease in $J_{\downarrow}$. At a
fixed value of $V_a$, the magnitude of $T_{\downarrow}(E_z)$
decreases so much that electrons with all possible values of $l$
give a very small contribution to the current, and it becomes
minimal. At $\sigma_z=1/2$ the current density $J_{\uparrow}$ is
only determined by electrons with $l=0$, so the contribution of
this current component to the total current density $J_t$ is
small. With a further increase in $V_a$, the second resonant peak
of $T_{\sigma_z}(E_z)$ intersects the line $E_{zm}^{\sigma_z}(0)$,
and a second peak appears in $J_{\downarrow}(V_a)$ and
$J_{\uparrow}(V_a)$. In this case, the width of the second peak of
$T_{\downarrow}(E_z)$ is so large that it intersects practically
all lines $E_{zm}^{\downarrow}(l)$ (at $V_a \geq 0.08$ V). As a
result, $J_{\downarrow}$ is produced by the electrons located at
all the filled Landau levels. For this reason the second peak of
$J_{\downarrow}(V_a)$ is higher and smoother than first one, and
it does not contain visible kinks. Note that the value of the
second peak of $J_{\uparrow}$ approximately equals to the value of
the first peak.

\begin{figure}
\mbox{
\includegraphics[width=16cm,height=16cm]{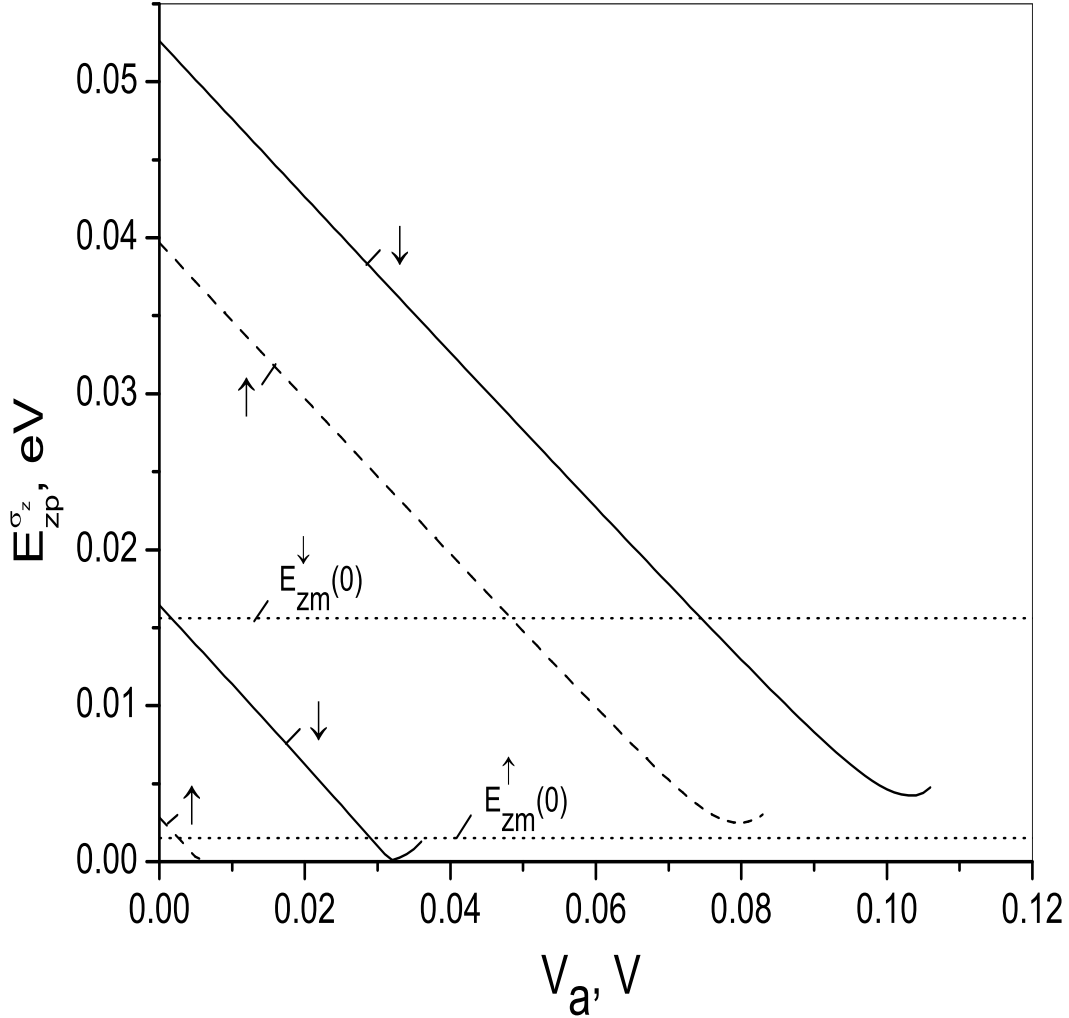}}
\vspace{-5mm} \caption{The bias-voltage dependence of the
 locations $E^{\sigma_z}_{zp}$ of two resonant peaks in the
dependencies $T_{\downarrow}(E_z)$ (solid lines) and
$T_{\uparrow}(E_z)$ (dashed lines) for $B=4$ T and $x_3=0.0$.}
\label{fig:6}
\end{figure}

Let us denote the resonant peak locations of $T_{\sigma_z}(E_z)$
by $E_{zp}^{\sigma_z}$. Fig.6 shows the dependencies of
$E_{zp}^{\sigma_z}(V_a)$ for the first two peaks of
$T_{\downarrow}(E_z)$ (solid lines) and $T_{\uparrow}(E_z)$
(dashed lines) for $B=4$~T, $E_F=10$~meV, and $x_3=0.0$. It is
seen from this figure that the first and second resonant peaks of
$T_{\downarrow}(E_z)$ are located in the region of smaller values
of $E_z$ than those of $T_{\uparrow}(E_z)$. With increasing $V_a$,
the locations of the resonant peaks of $T_{\sigma_z}(E_z)$ shift
to the low-energy region. Each current density component
$J_{\sigma_z}$ makes a contribution to the total current density
$J_{t}$ for those values of $V_a$ for which the value of
$E_{zp}^{\sigma_z}$ is less than the value of
$E_{zm}^{\sigma_z}(0)$. Note that the end-points of the
$E_{zp}^{\sigma_z}(V_a)$ dependencies correspond to the
disappearance of the resonant peaks in the $T_{\sigma_z}(E_z)$,
i.e. these dependencies are monotone with further increase of
$V_a$.

\subsection{The effect of total RTD current spin polarization}

It is clear that a high degree of the current spin polarization
occurs when $J_{\uparrow}$ is small. It follows from (\ref{eq8})
that at low temperatures the current is only created by the
electrons for which the condition $E_z < E_{zm}^{\sigma_z}(l)$ is
fulfilled. It is obvious that for the spin-up electrons
($\sigma_z=1/2$) located at the lowest Landau level ($l=0$), the
condition $E_{zm}^{\uparrow}(0) \leq 0$ can be fulfilled. This
implies that for the spin-up electrons, the lowest Landau level is
located higher than the Fermi level, and spin-up electrons are
absent in the RTD emitter. As a result, the effect of total spin
polarization of the electron current in the RTD must occur when
the current is only caused by the spin-down electrons
($J_{\uparrow}=0$, $P=1$).

\begin{figure}
\mbox{
\includegraphics[width=16cm,height=16cm]{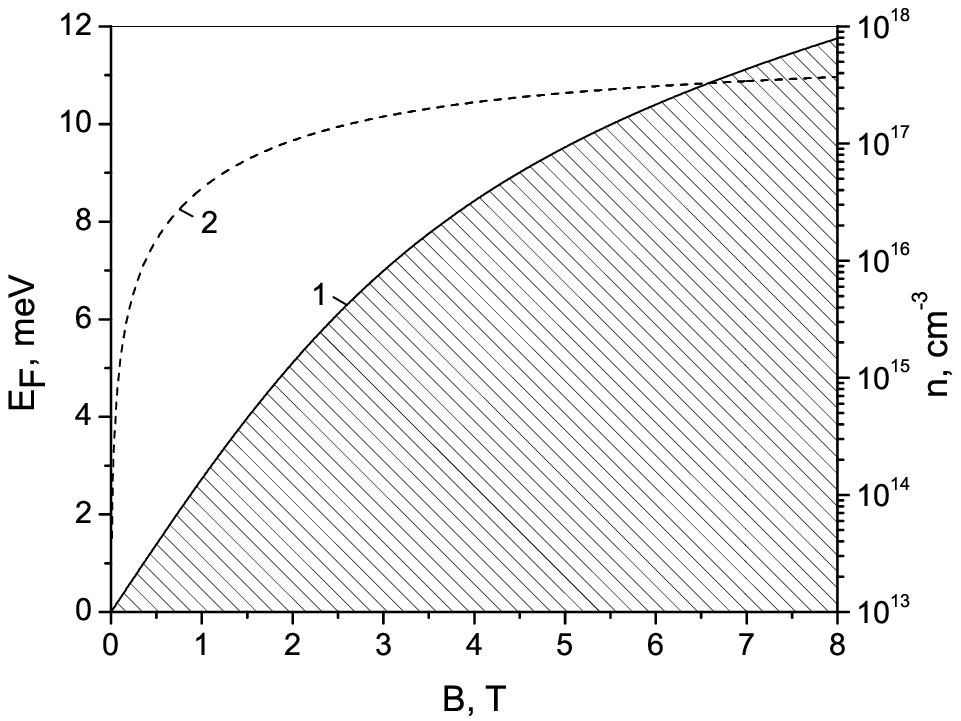}}
\vspace{-5mm} \caption{$E_F(B)$ (left ordinate axis) and $n(B)$
(right ordinate axis) corresponding to the occurrence of total
current spin polarization effect.} \label{fig:7}
\end{figure}

Let us show that the condition $E_{zm}^{\uparrow}(0) \leq 0$ can
be fulfilled for moderate magnetic fields $B$. In Fig.7 the
 $E_F(B)$ dependence (solid curve 1), corresponding to the solution
of equation $E_{zm}^{\uparrow}(0) = 0$, is plotted along the left
axis of the ordinates. The magnetic field dependence of the RTD
emitter electron concentration $n$ (dashed curve 2) is presented
along the right axis of the ordinates assuming that  $n$ is
related to $E_{F}$ by the equation
$n=(1/3\pi^2)(2m^*E_F/\hbar^2)^{3/2}$. (We consider the electron
gas in the RTD emitter to be degenerate).  For a fixed value of
$E_F$, the effect of the total spin polarization of the electron
current must occur starting at a critical value of $B$. (This
situation corresponds to the dashed area in Fig.7). Note that in
order to decrease the critical value of $B$, it is necessary to
decrease the value of $E_F$. For example, for the moderate value
$B=2$~T the effect of the total spin polarization of the electron
current occurs at $E_F=5.1$~meV. (The corresponding value of $n$
is $10^{17}$ cm$^{-3}$).

\begin{figure}
\mbox{
\includegraphics[width=16cm,height=18cm]{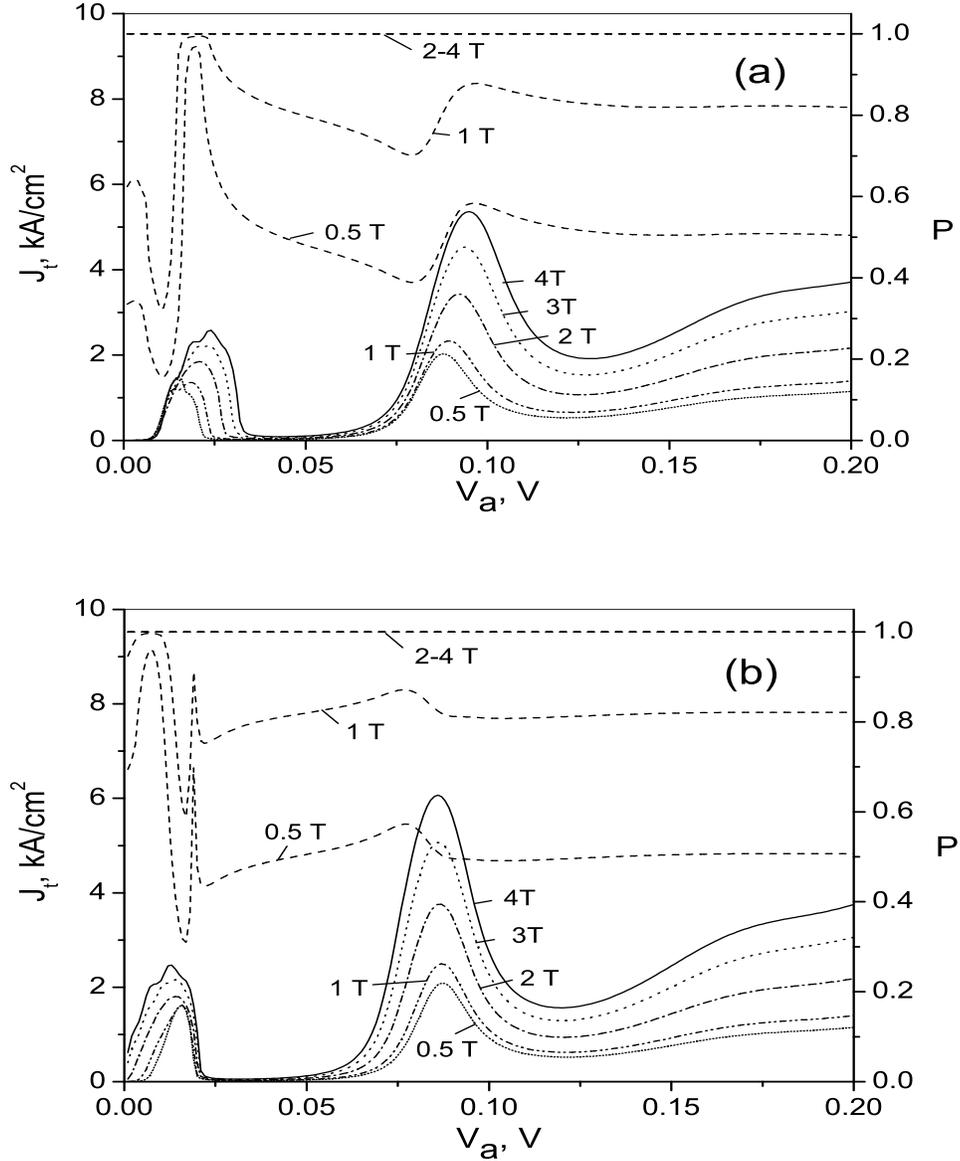}}
\vspace{-5mm} \caption{$J_t(V_a)$  (the left axis of ordinates)
and $P(V_a)$ (the right axis of ordinates)  for five values of
$B=0.5,1,2,3,4$~T, $E_F=5.1$~meV at (a) $x_3=0.0$ and (b)
$x_3=0.05$.} \label{fig:8}
\end{figure}

Now we consider the influence of constant magnetic field $B$ on
the $J_t(V_a)$ and $P(V_a)$ for two values of the Mn concentration
$x_3$ in the RTD quantum well at $E_F=5.1$~meV. In Fig.8
$J_t(V_a)$ (the left axis of ordinates, curves of different types
except dashed lines) and $P(V_a)$ (the right axis of ordinates,
dashed lines) are shown for (a) $x_3=0.0$, (b) $x_3=0.05$ for five
different values of $B=0.5,1,2,3,4$~T.  It is seen from Fig.8 that
with increasing $B$ the current density $J_t$ in the RTD
increases, and kinks on the first resonant peak of $J_t$ arise.
The value of $P$ also increases with increasing $B$. Starting with
$B=2$~T the electron current in the RTD is totally spin polarized
($P=1$). As one can see in Fig.8a, in the case $x_3=0.0$ the peaks
of the $J_t(V_a)$ coincide with the peaks of $P(V_a)$ . For the
case $x_3=0.05$ (Fig.8b) the situation is different because the
peak values of the current density correspond to the local minima
of $P$. So, we conclude that in moderately low magnetic fields
$B$, the maximal degree of the current spin polarization in the
peak values of the current takes place when the RTD quantum well
does not contain Mn ions. In this case the first current density
peak is characterized by almost total current spin polarization.

\begin{figure}
\mbox{
\includegraphics[width=16cm,height=18cm]{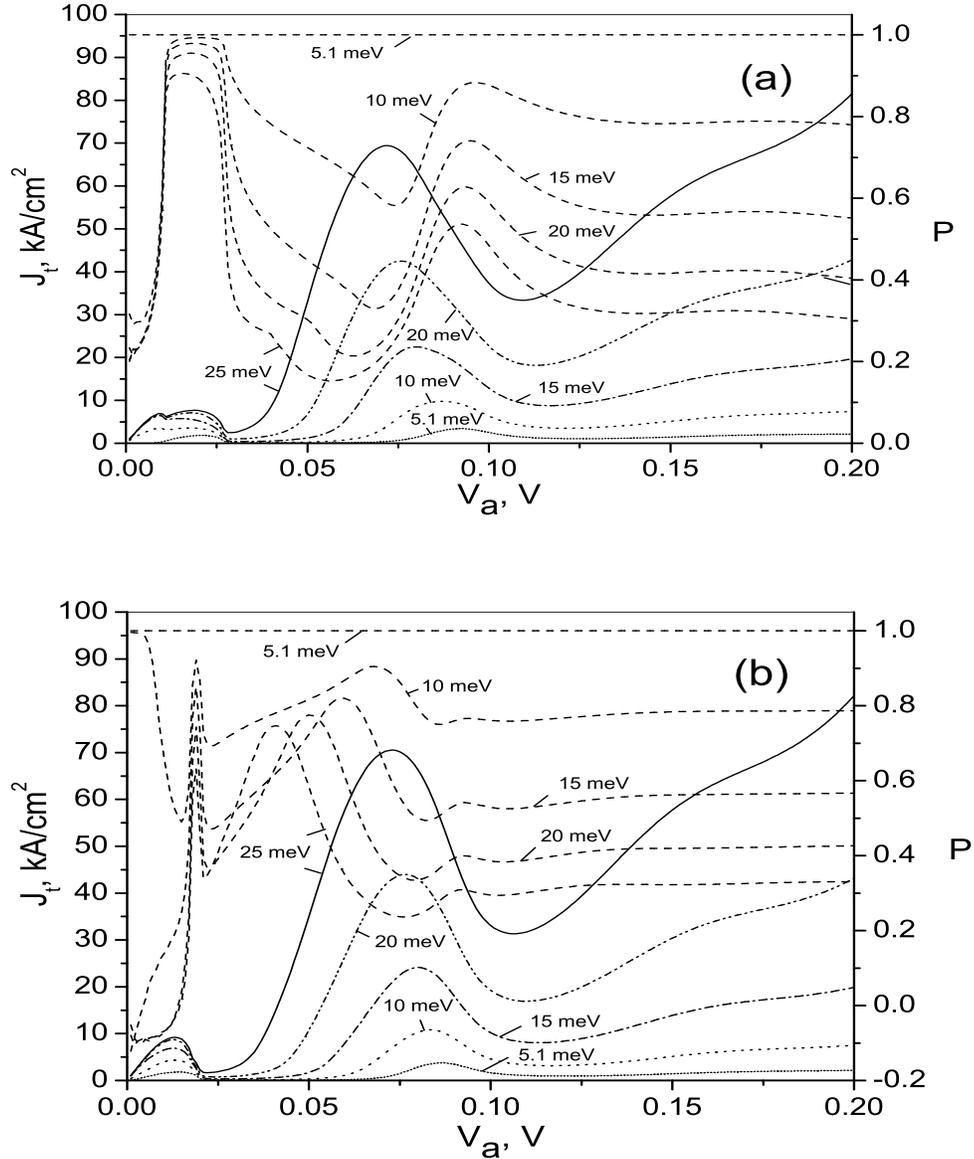}}
\vspace{-5mm} \caption{$J_t(V_a)$  (the left axis of ordinates)
and $P(V_a)$ (the right axis of ordinates)  for five values of
$E_F=5.1,1,2,3,4$~meV, $B=2$~T at (a) $x_3=0.0$ and (b)
$x_3=0.05$.} \label{fig:9}
\end{figure}

In order to obtain a high value of the spin-polarized current in
the RTD, it is necessary to increase $E_F$. Fig.9 shows $J_t(V_a)$
(the left axis of  ordinates, curves of different types except
dashed lines) and $P(V_a)$ (the right axis of ordinates, dashed
lines) at (a) $x_3=0.0$ and (b) $x_3=0.05$ for five different
values of $E_F=5.1,10,15,20,25$~meV at $B=2$~T.  It is seen from
Fig.9 that with increasing $E_F$ the current density peak values
increase. However, the value of $P$ decreases and, moreover, the
function $P$ becomes negative in low-voltage region for the case
$x_3=0.05$ (Fig.9b).  The first current density peak is
characterized by the high value of $P$ for the case $x_3=0.0$ as
usual, but the difference in $P$ for the second peak in cases
$x_3=0.0$ and $x_3=0.05$ becomes smaller. Note that the high value
of the peak-to-valley ratio typical of the first current density
peak also decreases  with increasing  $E_F$.

\section*{Conclusion}
In this paper, we have investigated theoretically the
spin-polarized electron current in a double-barrier semimagnetic
RTD based entirely on Zn$_{1-x}$Mn$_{x}$ Se semimagnetic
semiconductor. We have demonstrated the dependance of the current
spin polarization on the external constant magnetic field, the
applied voltage bias, and the distribution of Mn ions in the RTD.
We have obtained the condition for total current spin polarization
in the semimagnetic RTD, and we have found the optimal
distribution of Mn ions in the RTD providing the maximal current
spin polarization in the current peaks for arbitrary values of the
external magnetic fields and the Fermi levels in the RTD emitter.
We have demonstrated that the degree of current spin polarization
in the semimagnetic RTD can be effectively controlled by an
electric field, and this fact can be used for creating the voltage
controlled sources of spin polarized current for spintronics
devices.

\section*{Acknowledgments}

We are grateful to G. D. Doolen for useful discussions. This work
was supported by the Department of Energy (DOE) under Contract No.
W-7405-ENG-36, by the National Security Agency (NSA), and by the
Advanced Research and Development Activity (ARDA).



\newpage
 \begin{center}
Figure captions
 \end{center}
\begin{description}
\item [Fig.1] (a) Zn$_{1-x}$Mn$_{x}$Se double-barrier
resonant-tunnelling semimagnetic
 nanostructure (RTD)  and (b) its spin-dependent conduction band profile  at the nonzero bias voltage.

\item [Fig.2] Dependence of the conduction-band edges of the RTD
quantum well on the Mn ion concentration $x_3$ for spin-down
(solid lines) and spin-up (dashed lines) electrons for
$B=2,3,4$~T.

\item [Fig.3] The zero bias voltage RTD potential profile for
spin-down electrons (solid lines) and for spin-up electrons
(dashed lines) for (a) $x_3=0.0$ and (b) $x_3=0.05$ at $B=4$~T.

\item [Fig.4] $J_{\uparrow}(V_a)$, $J_{\downarrow}(V_a)$,
$J_t(V_a)$ (the left axis of ordinates) and $P(V_a)$ (the right
axis of ordinates) at (a) $B=2$~T and (b) $B=4$~T for $x_3=0.0$,
$E_F=10$~meV.

\item [Fig.5] (a) $T_{\downarrow}(E_z)$ and (b)
$T_{\uparrow}(E_z)$ for the different values of the voltage bias
$V_a$ at $B=4$ T and $x_3=0.0$ (the numbers next to the curves
show the corresponding values of $V_a$ in volts).

\item [Fig.6] The bias-voltage dependence of the
$E^{\sigma_z}_{zp}$ locations of two resonant peaks in the
dependencies $T_{\downarrow}(E_z)$ (solid lines) and
$T_{\uparrow}(E_z)$ (dashed lines) for $B=4$ T and $x_3=0.0$.

 \item [Fig.7] $E_F(B)$ (left ordinate axis)
and $n(B)$ (right ordinate axis) corresponding to occurrence of
the total current spin polarization effect.

\item [Fig.8] $J_t(V_a)$  (the left axis of ordinates) and
$P(V-a)$ (the right axis of ordinates)  for five values of
$B=0.5,1,2,3,4$~T, $E_F=5.1$~meV at (a) $x_3=0.0$ and (b)
$x_3=0.05$.

\item [Fig.9] $J_t(V_a)$  (the left axis of ordinates) and
$P(V-a)$ (the right axis of ordinates)  for five values of
$E_F=5.1,1,2,3,4$~meV, $B=2$~T at (a) $x_3=0.0$ and (b)
$x_3=0.05$.

\end{description}

\end{document}